\documentclass[twocolumn,showpacs,preprintnumbers,amsmath,amssymb,nofootinbib,scrartcl]{revtex4}
\input psfig.sty 
\usepackage[dvips]{color}
\usepackage{graphicx}
\usepackage{epsfig}

\usepackage{blindtext}
\begin{document}

\title{Probing the time dependence of dark energy}
\author{E. M. Barboza Jr.$^{1,2}$\footnote{E-mail: edesiobarboza@uern.br}}

\author{J. S. Alcaniz$^{1}$\footnote{E-mail: alcaniz@on.br}}

\address{$^{1}$Observat\'orio Nacional, 20921-400, Rio de Janeiro - RJ, Brasil}

\address{$^{2}$Departamento de F\'{\i}sica, Universidade do Estado do Rio Grande do Norte, 59610-210, Mossor\'o - RN, Brasil}

\date{\today}

\begin{abstract}

A new method to investigate a possible time-dependence of the dark energy equation of state $w$ is proposed. We apply this methodology to two of the most recent data sets  of type Ia supernova (Union2 and SDSS) and the baryon acoustic oscillation peak at $z = 0.35$. For some combinations of these data, we show that there is a clear departure from the standard $\Lambda$CDM model at intermediary redshifts, although a non-evolving dark energy component ($dw/dz = 0$) cannot be ruled out by these data. The approach developed here may be useful to probe a possible evolving dark energy component when applied to upcoming observational data.

\end{abstract}


\maketitle

\section{Introduction}

Approaches to modeling the dark energy are based either on a particular choice of its equation of state (EoS) parameter $w(z)$ or on modifications of gravity at very large scales. In the context of general relativistic models, at least three different ways may be followed in order to find the dark energy EoS from observations. The first and most direct one is to solve the scalar field equations for a particular theory, which clearly cannot provide a model-independent parameter space to be compared with the observational data. Other possibilities are either to build a functional form for $w(z)$ in terms of its current value $w_0$ and of its time-dependence $w'$ or to perform a parameter-free approach, such as binned EoS, decomposition into orthogonal basis and principal component analysis.

However, irrespective of the choice made, it seems clear nowadays that it is extremely difficult, if not impossible, to single out the best-fit dark energy model on observational data basis only. As well observed in Ref.~\cite{linde}, one may, instead of trying to find which dark energy scenario is correct, test which models can be ruled out by the available observations. This, of course, will not unveil the nature of the dark energy but may, with the current and upcoming observational data, diminish considerably the range of possibilities.

An interesting example in this direction involves two of the favorite candidates for dark energy, namely, the vacuum energy density ($\Lambda$) and a dynamical scalar field ($\Phi$). Among other things, what observationally may differ these two candidates for dark energy is that, in the former case, the EoS associated with $\Lambda$ is constant along the evolution of the Universe ($w = -1$) whereas in generic quintessence scenarios $w(z)$ is a function of time~\cite{quint}. Therefore, taking this important difference into account, one may conclude that if any observable deviation from a constant EoS is consistently found, this naturally poses a problem for any model based on this assumption, which includes our standard cosmological scenario.

\begin{figure*}
\centerline{\psfig{figure=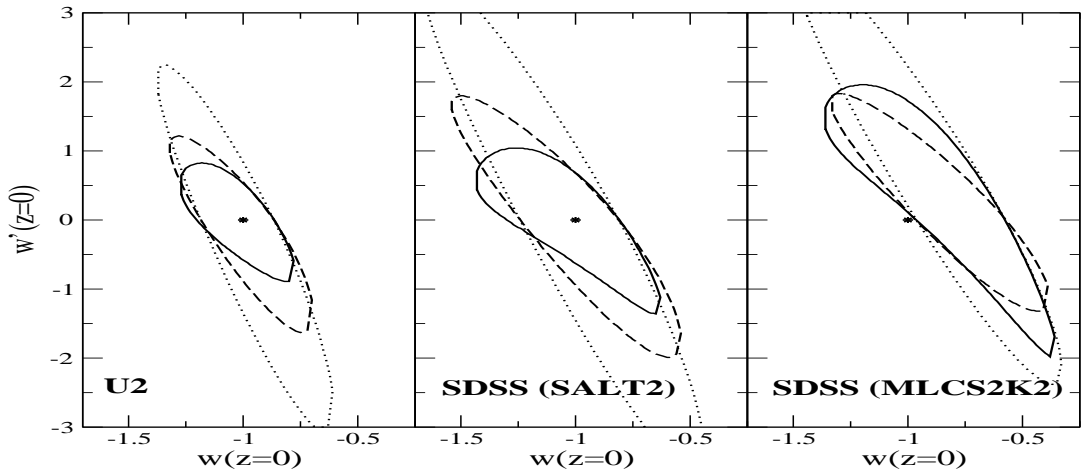,width=3.1truein,height=6.8truein,angle=-90}
\hskip 0.1in}
\caption{Contours of  $\Delta \chi^2 = 6.17$ in the plane $w(z^*=0) - w'(z^* = 0)$.  Solid contours stand for our second order expansion of $\rho_{\rm{DE}}(z)$ while dotted and dashed ones correspond, respectively, to CPL and the parameterization of Ref.~\cite{edesio}. Black dots indicate the $\Lambda$CDM model.}
\label{fig1}
\end{figure*}

In this paper, we propose a method to investigate a possible time-dependence of the dark energy equation of state from current observational data. Differently from other approaches that use parameterizations of $w(z)$~\cite{cpl,edesio,para} or $\rho_{\rm{DE}}(z)$~\cite{wf}, we use only the Taylor expansion of the dark energy density $\rho_{\rm{DE}}(z)$ around different values of $z$ and the conservation equation as a recurrence formula, so that the $\rho_{\rm{DE}}(z)$ derivatives can be directly related to $w(z)$ and its derivatives. In doing this, we reduce considerably the smearing effect (due to the multiple integrals that relate $w(z)$ to cosmological distances) that makes constraining $w(z)$ extremely difficult (see, e.g., \cite{maor}). We note that, unlike the first order parameterizations of $w(z)$, in which the time-dependence is determined uniquely by the value of $w'$ at a specific $z$ (usually $z = 0$), in this approach such variation may be verified at different values of $z$ by changing the expansion center $z^*$ at small redshift intervals. From this, we build $w(z)$ and $w'(z)$ from the data and discuss the behavior of the $w - z$ and $w' - z$ spaces.

We use in our analyses two of the most recent SNe Ia data sets, namely, the updated compilation of Union sample (U2)~\cite{union} and the  nearby + SDSS + ESSENCE + SNLS + Hubble Space Telescope (HST) set of 288 SNe Ia discussed in Ref.~\cite{sdss} (throughout this paper we refer to this set as SDSS compilation). We consider two sub-samples of this latter compilation that use SALT2~\cite{salt2} and MLCS2k2~\cite{mlcs2k2} SN Ia light-curve fitting methods. Along with the SNe Ia data, and to diminish the degeneracy between the two free parameters of the model we also use the baryonic acoustic oscillation (BAO) peak at $z_{\rm{BAO}} = 0.35$ of Ref.~\cite{bao}.

\section{The method}

Let us consider a Friedmann-Robertson-Walker universe dominated by non-relativistic matter (baryonic and dark) and some form of dark energy. The Friedmann equation for this scenario is given by
\begin{equation}
\label{eq.fried}
H^2\equiv\Big(\frac{\dot{a}}{a}\Big)^2=\frac{8\,\pi\,G}{3}\,\Big[\rho_{\rm{M}}(a)+\rho_{\rm{DE}}(a)\Big] -\frac{k}{a^2},
\end{equation}
where $H$ and $k$ stand for the Hubble and curvature parameters, $a = 1/(1+z)$ is the cosmological scale factor, and $\rho_{\rm{M}}$ and $\rho_{\rm{DE}}$ are, respectively, the non-relativistic matter and dark energy density parameters.

Now, let us suppose that $\rho_{\rm{DE}}(z)$ is an analytic function in the range $(z^*-\epsilon_{-},z^* +\epsilon_{+})$, so that its Taylor series,
\begin{eqnarray}
\label{eq.DEdensity}
\rho_{\rm{DE}}(z)&=&\rho_{\rm{DE}}(z^*)+\rho_{\rm{DE}}^{\prime}\Big{\vert}_{z^*}\,(z-z^*)+\\
&+&\frac{1}{2}\,\rho_{\rm{DE}}^{\prime\prime}\Big{\vert}_{z^*}\,(z-z^*)^2+\cdots, \nonumber 
\end{eqnarray}
converges to $\rho_{\rm{DE}}(z)$ in this range. Since the components of the cosmic fluid are separately conserved, the dark energy density $\rho_{\rm{DE}}(z)$ and its EoS parameter $w(z)$ are directly related by the conservation equation
\begin{equation}
\label{eq.cont}
\rho_{\rm{DE}}^{\prime}(z) = 3\,\rho_{\rm{DE}}(z)\,\frac{1+w(z)}{1+z}\;.
\end{equation}
Note that we can use the above equation as a recurrence formula to write the derivatives of $\rho_{\rm{DE}}$ at $z=z^*$ in terms of the EoS parameters derivatives, i.e.,
$$\begin{array}{lll}
\rho_{\rm{DE}}^{\prime}\Big{\vert}_{z^*}&=&\gamma\,\rho_{\rm{DE}}(z^*)\;,\\
&&\\
\rho_{\rm{DE}}^{\prime\prime}\Big{\vert}_{z^*}&=&\big[\,\gamma^2+\big(\,3\,w^{\prime}(z^*)-\gamma\,\big)/(1+z^*)\,\big]\rho_{\rm{DE}}(z^*)\;,\\
\  \vdots&&\  \vdots
\end{array}
$$
where $\gamma\equiv3\,[1+w(z^*)]/(1+z^*)$ and a prime denotes differentiation with respect to $z$. From the above results, we can rewrite the Friedmann equation (\ref{eq.fried}) as
\begin{equation}
\label{FriedmannEQ2}
{{H}}^2 = H_0^2\Big[\Omega_{\rm{M},0}(1+z)^3 + \frac{\Gamma}{\varphi_0}\varphi(z) + \Omega_{k,0}(1+z)^2\Big],
\end{equation}
where $\Gamma = 1-\Omega_{\rm{M},0}-\Omega_{k,0}$,
$$
\label{O2expoverr}
\varphi(z)=1+\gamma(z-z^*)+\frac{1}{2}\Big[\gamma^2+\frac{3w^{\prime}(z^*)-\gamma}{1+z^*}\Big](z-z^*)^2+\cdots\;, 
$$
and the subscript ``0" denotes present-day quantities. In what follows, we will restrict our analysis up to the second order expansion of $\rho_{\rm{DE}}(z)$. The practical reason for this choice is that it involves the minimal number of parameters [$w(z)$ and $w'(z)$] required to probe the EoS time-dependence. However, in doing so it is important to observe that we are also assuming that the real function describing the dark energy density is a well-behaved function of $z$ with a smooth behavior in the $z^* \pm \epsilon_{\pm}$ interval, so that the higher order terms are negligible compared to those considered. This assumption in turn is well justified since the result $w \approx \rm{const.}$ seems to be in good agreement with most of the current data~(see, e.g., \cite{sollerman}).

\begin{figure*}
\centerline{\psfig{figure=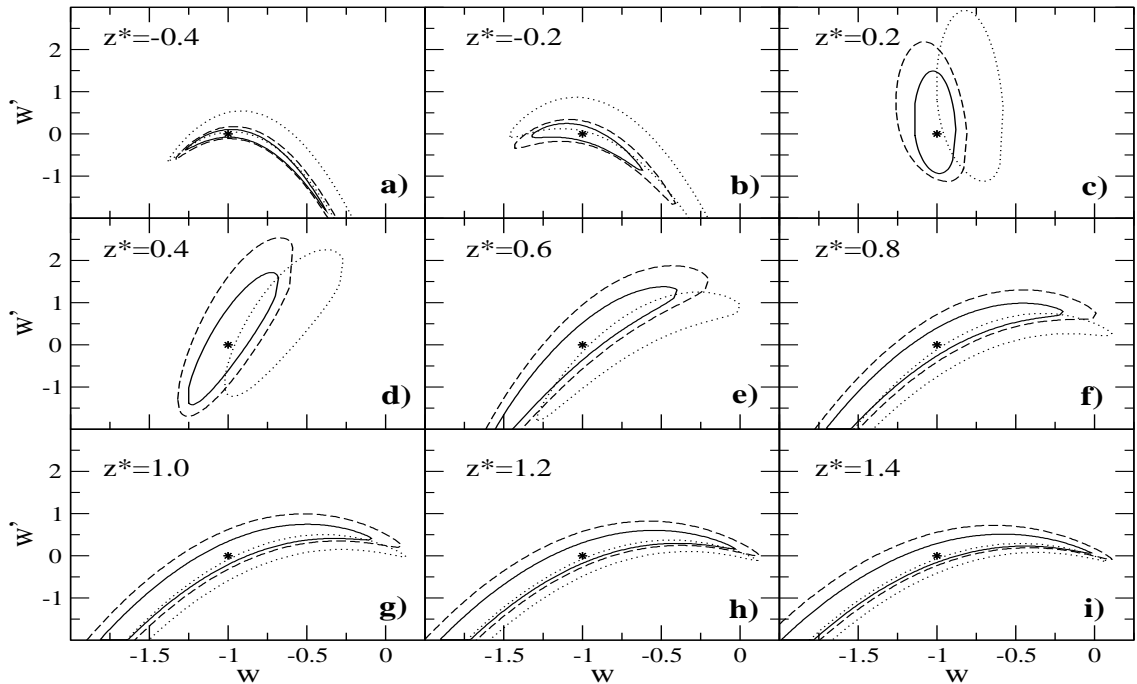,width=3.9truein,height=6.3truein,angle=-90}}
\caption{Contours of  $\Delta \chi^2 = 6.17$ in the plane $w - w'$ for different values of $z^*$. Here, solid, dashed and dotted  curves stand for the $\Delta \chi^2$ contours from U2 + BAO, SDSS (SALT2) + BAO and SDSS (MLCS2k2) + BAO, respectively.}
\label{fig1}
\end{figure*}

\section{Application, results and discussion}

In order to discuss the current observational constraints on $w(z)$ and $w'(z)$ we  use two of the most recent SNe Ia compilations available, namely, the Union2 sample of  Ref.~\cite{union} and the two compilations of the SDSS collaborations discussed in Ref.~\cite{sdss}. The Union2 sample\footnote{http://www.supernova.lbl.gov/} is an update of the original Union compilation. It comprises 557 data points including recent large samples from other surveys and uses SALT2 for SN Ia light-curve fitting. The SDSS compilation\footnote{http://www.sdss.org/} of 288 SNe Ia uses both SALT2 and MLCS2k2 light-curve fitters\footnote{It is worth emphasizing that SALT2 fitter does not provide a cosmology-independent distance estimate for each SN, since some parameters in the calibration process are determined in a simultaneous fit with cosmological parameters to the Hubble diagram (see~\cite{friemanON} for a discussion). See also \cite{sdss} for a discussion on possible systematic biases from the multicolor light curve shape method.} and is distributed in redshift interval $0.02 \leq z \leq 1.55$.

Along with the SNe Ia data, and to diminish the degeneracy between the dark energy parameters $\Omega_{\rm{M,0}}$, $w(z)$ and $w'(z)$, we also use the BAO parameter~\cite{bao}
\begin{equation}
{\cal{A}} = D_V{\sqrt{\Omega_{\rm{M,0}} H_0^2} \over {z_{\rm{BAO}}}}  = 0.469 \pm 0.017 \;,
\end{equation}
where $D_V = [r^2(z_{\rm{BAO}}){z_{\rm{BAO}}}/{H(z_{\rm{BAO}})}]^{1/3}$ is the so-called dilation scale, defined  in terms of the comoving distance $r$ to $z_{\rm{BAO}}$. The present value of the Hubble parameter $H_0$ is taken as a nuisance parameter and we marginalize over it.  In our statistical analysis, therefore, we minimize  the function $\chi^2 = \chi^{2}_{\rm{SNe}} + \chi^{2}_{\rm{BAO}}$, which takes into account all the data sets mentioned above. Motivated by the recent results of the CMB power spectrum~\cite{wmap}, we assume spatial flatness in the following analyses.

\begin{figure*}[t]
\centerline{\psfig{figure=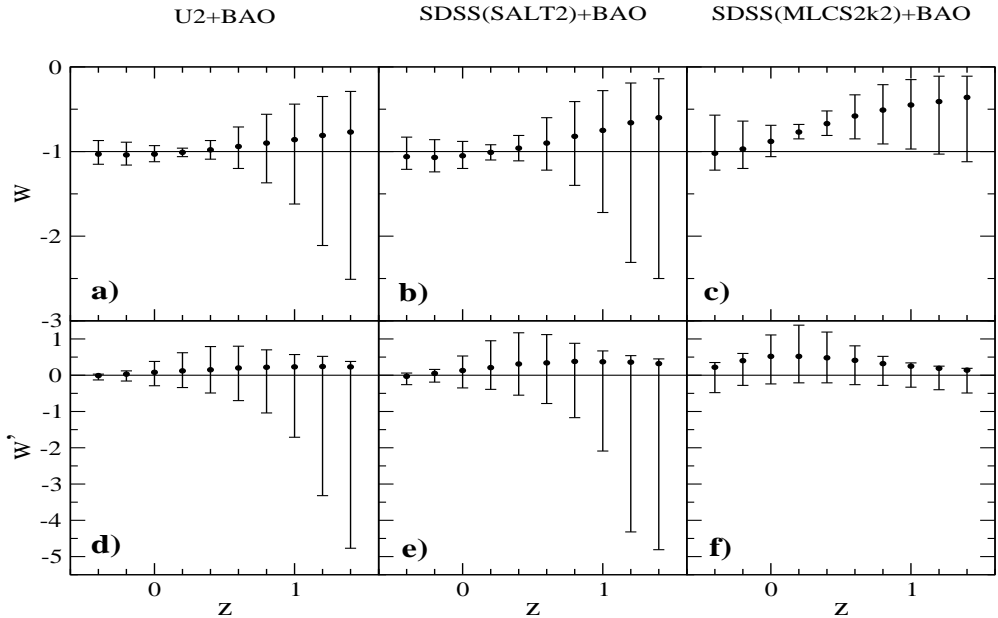,width=3.8truein,height=6.3truein,angle=-90}
\hskip 0.1in}
\caption{Evolution of $w$ and $w'$ arising from the combinations of data discussed in the text. Error bars correspond to $\Delta \chi^2 = 1$ ($1\sigma$). Clearly, no evidence for an evolving EoS is found over the entire redshift interval. Note, however, that  for SDSS (MLCS2k2) + BAO data a clear departure from the $\Lambda$CDM case ($w = -1$ and  $w' = 0$) is detected for some values of $z^*$.}
\label{fig1}
\end{figure*}

Figure 1 shows contours of $\Delta \chi^2 = 6.17$ arising from the joint analysis U2 + BAO (Panel 1a) , SDSS (SALT2) + BAO (Panel 1b)  and  SDSS (MLCS2k2)+BAO (Panel 1c) for the second order expansion around $z^*=0$ (solid contour). For the sake of comparison we also display the corresponding $\Delta \chi^2$ contours for two different first order parameterizations, i.e.,
\begin{equation}
w(z) = w_0 + {w_a}\frac{z}{1+z}\;,
\end{equation}
the so-called CPL parameterization~\cite{cpl} (dotted contour) and
\begin{equation}
w(z)=w_0+w_a\frac{z(1+z)}{1+z^2}\;,
\end{equation}
proposed in Ref.~\cite{edesio} (dashed contour)\footnote{Differently from CPL expression, this latter EoS parameterization is a bounded function of the redshift throughout the entire cosmic evolution, which allows to study the effects of a time varying EoS component to the distant future at $z \simeq -1$ as well as back to the last scattering surface of the CMB at $z \simeq 1100$ (see also~\cite{para} for other first order parameterizations).}. We note that, although there is space for both negative and positive values of $w'(z^* = 0)$ (or, equivalently, $w_a$), as well as for phantom-like~\cite{phantom} and quintessence-like~\cite{quint} dark energy,  the allowed parameter space for our second order expansion can be considerably reduced relative to the one produced by the first order EoS parameterizations\footnote{For a subset of model parameters, both parameterizations can be approximated by their second order expansion. As an example, by taking the best-fit values of $w_0$ and $w'_0$ from U2 sample (see Table I), we found that the relative error $\Delta\rho/\rho \lesssim 10\%$ at $z \simeq 1.5$, where  $\Delta\rho$ is the difference between the energy density derived from the above parameterizations and their second order expansions.}. We believe that this constraining power shown in these analyses  may be a consequence of the reduction of the smearing effect mentioned earlier~\cite{maor}, since in our approach the parameters $w(z^*)$ and $w'(z^*)$ appear in the expansion rate (Eq. (\ref{FriedmannEQ2})) not as an exponent of the $\rho_{\rm{DE}}$ function (as in the case of first order parameterizations). 

\begin{table*}

\begin{tabular}{|l|l|l|l|l|l|l|l|l|}
\hline
&\multicolumn{2}{l|}{\quad \quad \quad U2+BAO}&\multicolumn{2}{l|}{\quad SDSS(SALT2)+BAO}&\multicolumn{2}{l|}{SDSS(MLCS2k2) + BAO}\\
\cline{2-7}
\quad  $z$&\quad \quad $w$&\quad \quad$w'$&\quad \quad $w$&\quad \quad $w'$& \quad \quad$w$&\quad  \quad $w'$\\
\hline\hline
$-0.4$&$-1.03^{+0.16}_{-0.12}$&$-0.01^{+0.04}_{-0.12}$&$-1.06^{+0.23}_{-0.15}$&$-0.03^{+0.09}_{-0.23}$&$-1.02^{+0.45}_{-0.20}$&\quad$0.22^{+0.13}_{-0.70}$\\
$-0.2$&$-1.04^{+0.15}_{-0.12}$&\quad$0.03^{+0.09}_{-0.19}$&$-1.07^{+0.21}_{-0.17}$&\quad$0.05^{+0.11}_{-0.24}$&$-0.97^{+0.33}_{-0.23}$&\quad$0.40^{+0.20}_{-0.68}$\\
\quad$0.0$&$-1.03^{+0.10}_{-0.09}$&\quad$0.08^{+0.30}_{-0.37}$&$-1.05^{+0.17}_{-0.15}$&\quad$0.13^{+0.40}_{-0.48}$&$-0.88^{+0.19}_{-0.18}$&\quad$0.52^{+0.59}_{-0.76}$\\
\quad$0.2$&$-1.01^{+0.05}_{-0.05}$&\quad$0.12^{+0.50}_{-0.46}$&$-1.01^{+0.09}_{-0.09}$&\quad$0.21^{+0.74}_{-0.60}$&$-0.77^{+0.09}_{-0.08}$&\quad$0.52^{+0.86}_{-0.73}$\\
\quad$0.4$&$-0.98^{+0.11}_{-0.11}$&\quad$0.15^{+0.64}_{-0.64}$&$-0.96^{+0.15}_{-0.15}$&\quad$0.31^{+0.86}_{-0.86}$&$-0.67^{+0.15}_{-0.14}$&\quad$0.48^{+0.71}_{-0.69}$\\
\quad$0.6$&$-0.94^{+0.23}_{-0.26}$&\quad$0.20^{+0.60}_{-0.90}$&$-0.90^{+0.30}_{-0.32}$&\quad$0.34^{+0.78}_{-1.12}$&$-0.58^{+0.25}_{-0.27}$&\quad$0.41^{+0.40}_{-0.67}$\\
\quad$0.8$&$-0.90^{+0.34}_{-0.47}$&\quad$0.22^{+0.48}_{-1.26}$&$-0.82^{+0.41}_{-0.58}$&\quad$0.38^{+0.50}_{-1.55}$&$-0.51^{+0.30}_{-0.40}$&\quad$0.32^{+0.20}_{-0.60}$\\
\quad$1.0$&$-0.82^{+0.42}_{-0.76}$&\quad$0.23^{+0.86}_{-0.34}$&$-0.75^{+1.94}_{-0.97}$&\quad$0.37^{+0.30}_{-2.46}$&$-0.45^{+0.30}_{-0.52}$&\quad$0.25^{+0.09}_{-0.58}$\\
\quad$1.2$&$-0.81^{+0.46}_{-1.30}$&\quad$0.24^{+0.28}_{-3.56}$&$-0.66^{+0.47}_{-1.65}$&\quad$0.36^{+0.18}_{-4.68}$&$-0.41^{+0.30}_{-0.62}$&\quad$0.19^{+0.06}_{-0.59}$\\
\quad$1.4$&$-0.77^{+0.48}_{-1.74}$&\quad$0.23^{+0.15}_{-5.00}$&$-0.60^{+0.46}_{-1.90}$&\quad$0.32^{+0.13}_{-5.13}$&$-0.38^{+0.25}_{-0.76}$&\quad$0.14^{+0.05}_{-0.63}$\\
\hline
\end{tabular}
\caption{Results for $w$ and $w'$ for the combinations of data discussed in the text. The error bars correspond to $1\sigma$ ($\Delta \chi^2 = 1$).}
\end{table*}

However, as mentioned earlier, the most interesting aspect in the above approach is that a possible time evolution of $w(z)$ and of its derivatives can be verified from the data at different values of $z^*$ by changing the expansion center at small redshift intervals. This procedure differs fundamentally from usual first order parameterizations in that variations of the EoS parameter at $z \neq 0$ cannot be detected if $w(z)$ is a smooth function around $z \simeq 0$. In other words, this amounts to saying that in the approach discussed in this paper, finding the pair $(w(z^* =0),w^{\prime}(z^* =0))=(-1,0)$ inside a significant confidence region cannot be taken as a definitive argument for a non-evolving dark energy since a single value of $w^{\prime} \neq 0$ at any $z^* \neq 0$ is enough to ensure its time-dependence.

This possibility is tested in Figures 2a-2i, where we show contours of $\Delta \chi^2 = 6.17$ in the $w$ - $w'$ plane for the three combinations of data discussed above. The contours are obtained as follows:

\begin{enumerate}

\item fix the expansion center $z^*$ in Eq. (\ref{FriedmannEQ2});

\item perform the statistical analysis described above to obtain the values of $\Omega_{\rm{M,0}}$, $w$ and $w^{\prime}$ and their corresponding error bars at $z^*$;

\item set $z^*\to z^*+\Delta\,z^*$ and repeat the previous step to estimate the above parameters at $z^*+\Delta\,z^*$.

\end{enumerate}


In our analysis, we divided the entire interval in $\Delta z^* = 0.2$ from $z = 0$ up to the highest redshift of our sample, i.e., $z = 1.4$. In Panels 2a-2i, solid, dashed and dotted  curves stand for the $\Delta \chi^2$ contours from U2 + BAO, SDSS (SALT2) + BAO and SDSS (MLCS2k2) + BAO, respectively.  For all combinations of SNe Ia + BAO, it is clear  that the second order expansion is fully compatible with a non-evolving EoS ($w' = 0$)  for the entire redshift interval considered. The same, however, cannot be said when we consider specifically the standard $\Lambda$CDM case and the combination SDSS (MLCS2k2) + BAO (dotted line). In this case, $\rho_{\rm{DE}} = \rm{const.}$ or, equivalently, $w = -1$ and $w' = 0$ (see Eq.~\ref{eq.DEdensity}), which is off from the best-fit value by $\gtrsim 2\sigma$ (c.l.) for several values of $z^*$. For the sake of completeness, we also show the resulting contours for two analyses with $z^* < 0$, i.e., $z^* = -0.4$ (Panel 2a) and $z^* = -0.2$ (Panel 2b).

Figure 3 shows the quantities $w$ and $w'$ as functions of the redshift for the combination of data discussed above. The error bars correspond to the confidence intervals obtained for $\Delta \chi^2 = 1$. Following the curve indicated by the best-fit points, the $w - z$ plane for all combinations seems to indicate a freezing behavior, which in agreement with recent discussions using reconstructing techniques such as the principal component analysis~\cite{krauss} and the maximum entropy method~\cite{trotta}. From these Panels, we can see more clearly that, although compatible with a time-independent EoS parameter ($w' = 0$), some combinations of the current data (SDSS (MLCS2k2) + BAO) can show a departure from the $\Lambda$CDM case ($w = -1$ and  $w' = 0$) for some values of $z^*$. More quantitatively, this can be seen in Table I, where we display the numerical values obtained in our analyses. For completeness, we also tested the influence of a nonvanishing curvature parameter on these results (see, e.g.,~\cite{clarkson} for a discussion). By considering the current interval from WMAP + $H_0$ + BAO data, i.e., $\Omega_k = -0.0023^{+0.0054}_{-0.0056}$~\cite{wmap}, we verified that all the above conclusions remain unchanged.

\section{Conclusions}

We have proposed a new method to constrain a possible time-dependence of the dark energy from current observational data. By expanding the function $\rho_{\rm{DE}}(z)$ around different values of $z$ and using the conservation equation as a recurrence formula, we have shown a clear compatibility of current SNe and BAO data with a non-evolving EoS for all samples considered. Our method is based on the  assumption that the dark energy density has a smooth behavior up to $z \sim 1.5$, so that higher order terms in the $\rho_{\rm{DE}}$ expansion are negligible when compared with the first and second ones. Besides involving  the minimal number of parameters $w(z)$ and $w'(z)$ required to probe the EoS time-dependence (see Eq.~\ref{FriedmannEQ2}), this seems to be a plausible assumption since a dark energy behavior somewhat close to a cosmological constant ($\rho_{\rm{DE}} = \mbox{const.}$) is fully supported by many analyses (see, e.g., \cite{pad} and refs. therein). We have also discussed some advantages of this approach over first order parameterizations commonly used in the literature and shown that the so-called smearing effect in the determinations of the $w(z)$ evolution may be greatly reduced. We believe that the next generation of experiments dedicated to measure the dark energy properties (mainly those measuring the expansion history from high-$z$ SNe Ia, baryon oscillations, etc. -- see, e.g., \cite{projects}) will probe cosmology with sufficient accuracy to decide between an evolving and a non-evolving dark energy EoS. In this regard, the methodology discussed here can be useful.

\begin{acknowledgments}
The authors are very grateful to Saulo Carneiro and Jackson Maia for a critical reading of the manuscript. EMBJr and JSA are supported by CNPq - Brazil.
\end{acknowledgments}

\end{document}